\documentclass[fleqn,twoside,twocolumn,nofootinbib,showkeys]{revtex4} 
\usepackage[sec,nocpr]{ujp} 

\begin{document}
\title[Influence of Surface Centers ]
{INFLUENCE OF SURFACE CENTERS\\ ON THE EFFECTIVE SURFACE
RECOMBINATION RATE AND
THE PARAMETERS OF SILICON SOLAR CELLS}%
\author{V.P.~Kostylyov}
\affiliation{V.E. Lashkaryov Institute of Semiconductor Physics, Nat. Acad. of Sci. of Ukraine}
\address{41, Prosp. Nauky, Kyiv 03680, Ukraine}
\email{sach@isp.kiev.ua}
\author{A.V.~Sachenko}
\affiliation{V.E. Lashkaryov Institute of Semiconductor Physics, Nat. Acad. of Sci. of Ukraine}
\address{41, Prosp. Nauky, Kyiv 03680, Ukraine}
\email{sach@isp.kiev.ua}
\author{I.O.~Sokolovskyi}
\affiliation{V.E. Lashkaryov Institute of Semiconductor Physics, Nat. Acad. of Sci. of Ukraine}
\address{41, Prosp. Nauky, Kyiv 03680, Ukraine}
\email{sach@isp.kiev.ua}
\author{V.V.~Chernenko}
\affiliation{V.E. Lashkaryov Institute of Semiconductor Physics, Nat. Acad. of Sci. of Ukraine}
\address{41, Prosp. Nauky, Kyiv 03680, Ukraine}
\email{sach@isp.kiev.ua}
\author{T.V.~Slusar}
\affiliation{V.E. Lashkaryov Institute of Semiconductor Physics, Nat. Acad. of Sci. of Ukraine}
\address{41, Prosp. Nauky, Kyiv 03680, Ukraine}
\email{sach@isp.kiev.ua}
\author{A.V.~Sushyi}
\affiliation{V.E. Lashkaryov Institute of Semiconductor Physics, Nat. Acad. of Sci. of Ukraine}
\address{41, Prosp. Nauky, Kyiv 03680, Ukraine}
\email{sach@isp.kiev.ua}

\udk{621.315.592} \pacs{71.55.Cn, 72.20.Jv,\\[-3pt] 72.40.+w}
\razd{\secvii}

\autorcol{V.P.\hspace*{0.7mm}Kostylyov, A.V.\hspace*{0.7mm}Sachenko,
I.O.\hspace*{0.7mm}Sokolovskyi et al.}

\setcounter{page}{362}%

\begin{abstract}
The results of our researches of the influence of
exponentially distributed surface centers on the effective surface
recombination rate and the parameters of silicon solar cells (SCs)
are reported. In our calculations, we assumed the acceptor and donor
surface states to lie in the upper and lower, respectively, parts of
the bandgap. The model also supposed a discrete surface level to
exist in the middle of the energy gap. In the case where the
integrated concentration of continuously distributed centers is
comparable with that of deep surface levels, those centers can affect
the SC parameters only due to the recombination. If the concentration of
continuously distributed centers is comparable or higher than the
concentration characterizing a charge built-in into the insulator,
those centers directly affect the surface band bending and the
photo-induced electromotive force. With the help of a computer
simulation, the conditions for the rate of surface recombination through
continuously distributed surface centers to exceed that through the
deep discrete level are determined. A decrease of the open-circuit
voltage in inverted silicon SCs associated with the recombination
through continuously distributed centers is calculated. The
obtained theoretical results are compared with the experimental
data.
\end{abstract}

\maketitle

\section{Introduction}

Nowadays, researchers deal with a quite considerable number of
semiconductor interfaces, the properties of which are substantially
governed by surface electron states. Such objects include, in particular,
the interface between a semiconductor and an insulator. The charge
characteristics of those interfaces have been studied more
completely than the recombination ones. While studying the
former, it was found that the acceptor surface states are located,
as a rule, in the upper part of the energy gap and the donor surface
states in the lower one. Depending on the energy, the surface states
with either a peak-like distribution or with the exponential
dependence of their density on the energy--the so-called
continuously distributed surface states--can be realized.
Information concerning the charge characteristics of the Si--SiO$_{2}$
interface was summarized, e.g., in monograph \cite{1}. In
particular, it was shown in work \cite{1} that the consideration of the
exponential distribution of surface centers in the silicon bandgap
into account allows one to explain the experimental results obtained, while studying
the generation parameters of silicon-based
metal--insulator--semiconductor structures.\looseness=1

As to the mechanism of surface recombination formation, it is supposed in
modern works on this subject that a discrete surface level located near the
middle of the energy gap makes a crucial contribution to the recombination. The
same can be said about works devoted to the analysis of the surface
recombination in solar cells on the basis of single-crystalline silicon
(see, e.g., works \cite{2,3}, in which either silicon dioxide or aluminum
oxide was used to reduce the surface recombination rate).

Researches of such promising versions of silicon solar cells as elements
with rear-side metallization have been activated recently \cite{4,5,6}. In
those elements, the task to minimize the rate of surface recombination on
the illuminated surface is extremely challenging.

This work is aimed at analyzing the relation between the contributions made to
the surface recombination by surface levels located near to and far from the
middle of the energy gap, as well as its dependence on the Fermi level
position at the surface. It is shown that, under certain conditions, the
contribution of remote surface levels to the recombination can substantially
exceed the deep-level contribution. The influence of exponentially
distributed surface states on the open-circuit voltage and the short circuit
current in silicon-based solar cells has been analyzed.

\section{Formulation of the Problem}

According to work \cite{1}, let us consider the Si--SiO$_{2}$ interface and
suppose that the acceptor surface levels are located in the upper half of
the silicon energy gap, the donor surface levels are in the lower half, and their
concentrations exponentially grow, while approaching the corresponding band
edges; in addition, there is a discrete surface level near the energy gap
middle-point. Let us analyze their contributions to the surface recombination
taking into account that their cross-sections and, accordingly, the
coefficients of electron and hole capture change with the energy, as was
described in work \cite{1}. More specifically, the cross-section of the hole
capture by acceptor states grows, whereas the same parameter for electrons
decreases, as the distance from the middle of the energy gap increases, with
the energy dependence for the cross-sections being steeper than that for the
center concentration. For donor levels located in the lower half of the
energy gap, the cross-sections of the electron capture grow, whereas those of
the hole capture decrease. For the sake of simplicity, let the system of centers
be symmetric. In the middle of the energy gap, there is a discrete surface
level with concentration $N_{t}$. In addition, the concentrations of donor
and acceptor centers per unit energy interval and the coefficients of
electron and hole capture by continuously arranged surface levels behave themselves as
follows:
\[
N_{a}(E)=N_{0a}\exp(E/E_{0}),
\]\vspace*{-7mm}
\[
N_{d}(E)=N_{0d}\exp(-E/E_{0}),
\]\vspace*{-7mm}
\[
C_{na}(E)=C_{0}\exp(-E/E_{r}),
\]\vspace*{-7mm}
\[
C_{nd}(E)=C_{0}\exp(-E/E_{r}),
\]\vspace*{-7mm}
\[
C_{pa}(E)=C_{0}\exp(E/E_{r}),
\]\vspace*{-7mm}
\begin{equation}
C_{pd}(E)=C_{0}\exp(E/E_{r}).
\end{equation}%
Here, $N_{0a}$ and $N_{0d}$ are the concentrations of donor and acceptor
centers, respectively, per unit energy interval at the energy-gap middle
point; $C_{0}$ is the capture factor at $E=0$; and $E_{0}$ and $E_{r}$ are
the characteristic energies of variations in the concentration and the capture cross-section,
respectively.

From the electroneutrality equation for the system with
regard for the recharging of surface levels at the illumination, the
dimensionless nonequilibrium band bending $y_{s}$ can be determined.
Let us include the
charge built-in into the insulator,
$N_{\mathrm{ins}}=Q_{\mathrm{ins}}/q,$ in the electroneutrality equation. Then, in the case of an
$n$-semiconductor, the equation looks like
\begin{equation*}
-\int\limits_{0}^{E_{g}/2}N_{0a}\exp(E/E_{0})f_{a}(E)dE-N_{t}f(E_{t})-N_{\mathrm{ins}}\,+
\end{equation*}\vspace*{-5mm}
\begin{equation*}
+\int\limits_{-E_{g}/2}^{0}N_{0d}\exp(-E/E_{0})(1-f_{d}(E))dE\,+
\end{equation*}\vspace*{-7mm}
\begin{equation}
+\,2N_{d}L_{D}\sqrt{-y_{s}}+(p_{0}+\Delta
p)\int\limits_{0}^{w}\exp(-y)dx=0.
\end{equation}
In this expression for donor and acceptor centers,
\begin{widetext}\vspace*{-7mm}
\begin{equation*}
f(E)=\frac{C_{n}(E)n(0)+C_{p}(E)n_{i}\exp(-E/kT)}{C_{n}(E)(n(0)+n_{i}\exp(\frac{E}{%
kT}))+C_{p}(E)(p(0)+n_{i}\exp(-\frac{E}{kT}))}
\end{equation*}%
\end{widetext}
is the nonequilibrium Fermi distribution for the surface level with
energy $E$ and the capture factors $C_{n}(E)$ and $C_{p}(E)$ for
electrons and holes, respectively; $n_{0}$ and $p_{0}$ are the
equilibrium concentrations of electrons and holes, respectively, in
the bulk; and $n(0)$ and $p(0)$ are the same quantities, but at the
semiconductor surface. We also use the following notation: $n_{i}$
is the concentration of charge carriers in the intrinsic
semiconductor, $E_{g}$ the energy gap width in the semiconductor,
$L_{D}$ the Debye screening length, $N_{d}$ the bulk concentration
of completely ionized donors, $w$ the thickness of the space charge
region (SCR) in the semiconductor, $k$ the Boltzmann constant, and
$T$ the temperature. The energy $E$ is reckoned from the middle of
the energy gap at the semiconductor surface.

In order to find a relation between the surface and bulk
concentrations of nonequilibrium charge carriers, let us use the
approximation of constant Fermi quasilevels for electrons and holes
in the SCR. In this case, for the injection level $\Delta
p$,\vspace*{-0.5mm}
\begin{equation}
n(0)=(n_{0}+\Delta p)e^{y_{s}},
\end{equation}\vspace*{-9mm}
\begin{equation}
p(0)=(p_{0}+\Delta p)e^{-y_{s}}.
\end{equation}

In Fig.~1, the theoretical dependences of the surface photovoltage
(in millivolt units) on the injection level calculated with the use
of Eqs.~(1)--(4) are depicted. The calculations were carried out for
\mbox{$N_{\rm ins}=10^{12}~\mathrm{cm}^{-2}$}, and the varied
calculation parameter was the concentration of exponentially
distributed surface levels $N_{0d}$ (it has the dimensionality$\ $of
$\mathrm{cm}^{\mathrm{2}}$/eV). The highest photovoltage values were
obtained in the cases $N_{0d}=0$ (curve~\textit{1}) and
$\int_{0}^{E_{g}/2}N_{0d}\exp (E/E_{0})dE\ll N_{\mathrm{ins}}$
(curve~\textit{2}). If the integrated concentration of exponentially
distributed surface centers is comparable with $N_{\mathrm{ins}}$,
the photovoltage magnitude decreases, and the strongest decrease
takes place at large enough injection levels (curves~\textit{3} and
\textit{4}). This is a manifestation of the effect when a charge
built-in into an insulator is screened by the charge of surface
centers.

With regard for Eqs.~(3) and (4), the densities of surface
recombination fluxes, i.e. the density of recombination currents
normalized by the electron charge $q$ that flow through the discrete
surface level and through the continuously distributed acceptor and
donor levels are equal, \mbox{respectively, to}\looseness=1
\begin{widetext}\vspace*{-5mm}
\begin{equation}
J_{rt}=\frac{C_{nt}C_{pt}N_{t}((n_{0}+\Delta p)(p_{0}+\Delta
p)-n_{i}^{2})}{C_{nt}(n(0)+n_{i}\exp(E_{t}/kT))+C_{pt}(p(0)+n_{i}\exp(-E_{t}/kT))},
\end{equation}\vspace*{-5mm}
\begin{equation}
J_{ra}=\int\limits_{0}^{E_{g}/2}\frac{C_{na}(E)C_{pa}(E)N_{0a}\exp(E/E_{0})((n_{0}+%
\Delta p)(p_{0}+\Delta
p)-n_{i}^{2})}{C_{na}(E)(n(0)+n_{i}\exp(E/kT))+C_{pa}(E)(p(0)+n_{i}\exp(-E/kT))}dE,
\end{equation}\vspace*{-5mm}
\begin{equation}
J_{rd}=\int\limits_{-E_{g}/2}^{0}\frac{%
C_{nd}(E)C_{pd}(E)N_{0d}\exp(-E/E_{0})((n_{0}+\Delta p)(p_{0}+\Delta
p)-n_{i}^{2})}{C_{nd}(E)(n(0)+n_{i}\exp(E/kT))+C_{pd}(E)(p(0)+n_{i}\exp(-E/kT))%
}dE.
\end{equation}%
\end{widetext}

\noindent The surface density of the total recombination flux is
determined by the expression
\begin{equation} J_{r}=J_{rt}+J_{ra}+J_{rd}.
\end{equation}%
Knowing the recombination fluxes, we can find the effective rates of surface
recombination through the corresponding levels,
\[
S_{t}=J_{rt}/\Delta p(U),\quad S_{a}=J_{ra}/\Delta p(U),
\]
\begin{equation}
S_{d}=J_{rd}/\Delta p(U)
\end{equation}
as well as the total effective rate of surface recombination,
\begin{equation}
S_{s}=S_{t}+S_{a}+S_{d}.
\end{equation}%
One can see from expressions (5)--(10) that the effective rate of surface
recombination through the deep discrete level can be determined using a
standard procedure. At the same time, to find the effective rates of surface
recombination through continuously distributed centers, it is necessary to
integrate the recombination fluxes through the acceptor and donor levels
over the energy.

\begin{figure}
\vskip1mm
\includegraphics[width=\column]{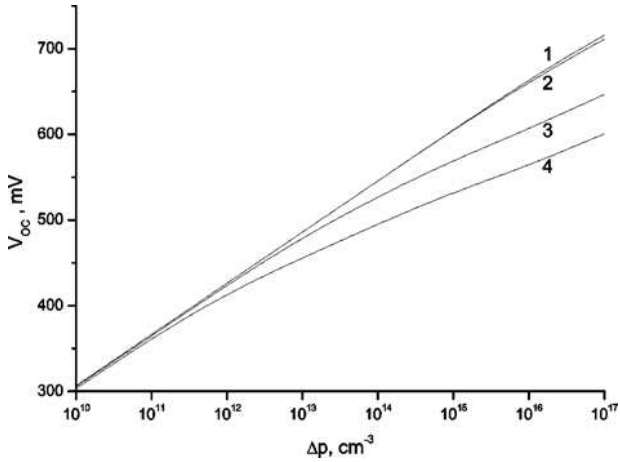}
\vskip-3mm\caption{Theoretical dependences of the surface
photovoltage on the injection level. The calculation parameters are
$N_{\mathrm{ins}}=$ $=10^{12}$~cm$^{-2}$, $n_{0}=10^{15}$~cm$^{-3}$,
and $N_{0d}kT=0$ (\textit{1}), $2\times10^{9}$ (\textit{2}),
$1.5\times10^{10}$ (\textit{3}), and
$2\times10^{10}~\mathrm{cm}^{-2}$ (\textit{4})
 }
\end{figure}

\section{Discussion of Results}

In Figs.~2 and 3, the dependences of the total effective surface
recombination rate $S_{s}$ and the effective recombination rates of
surface recombination through the deep level, $S_{t}$, and the
acceptor, $S_{a}$, and donor, $S_{d}$, states on the quantity
$\Delta p$ calculated for the cases $N_{\rm ins}=0$ and
$10^{12}~\mathrm{cm}^{-2}$, respectively, are depicted. The
following parameters were used at calculations: $N_{0a}kT=2\times
10^{9}~\mathrm{cm}^{-2}$, $N_{t}=10^{11}$~cm$^{-2}$,
$C_{nt}=C_{pt}=$ =~$10^{-9}$~cm$^{3}$/s, $E_{t}=0$,
$C_{0}=10^{-9}$~cm$^{3}$/s, $E_{0}=$ $=7.7kT$, $E_{r}=2.25kT$, and
$T=300~\mathrm{K}$. Note that the values taken for $E_{0}$ and
$E_{r}$ correspond to those quoted in work \cite{1}. One can see
that, for the selected $E_{0}$ and $E_{r}$ values, the
concentrations of continuously distributed levels grow more slowly
than the coefficients of hole capture by acceptor levels and
electron capture by donor levels. It should be noted that, at the
used $N_{t}$ and $N_{\mathrm{ins}}$ values and $N_{0a}kT=N_{0d}kT=$
$=2\times 10^{9}~\mathrm{cm}^{-2}$, the presence of surface levels
weakly affect the magnitude of $y_{s}$ (see curve~\textit{2} in
Fig.~1).\looseness=-1

\begin{figure}
\vskip1mm
\includegraphics[width=\column]{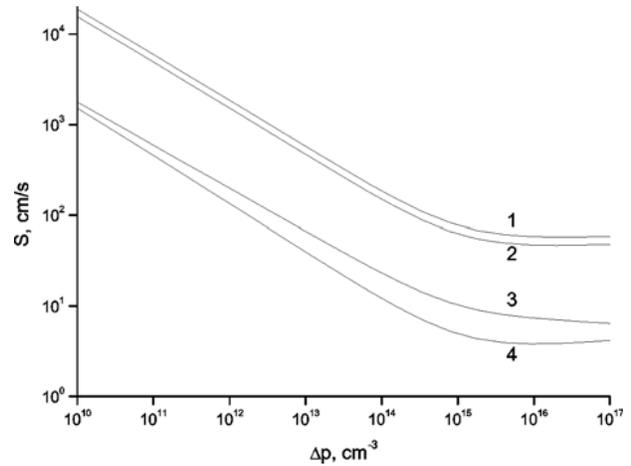}
\vskip-3mm\caption{ Theoretical dependences of the effective surface
recombination rates $S_{s}$ (\textit{1}), $S_{t}$ (\textit{2}),
$S_{d}$ (\textit{3}), and $S_{a}$ (\textit{4}) on the injection
level. The calculation parameters are $N_{\mathrm{ins}}=0$,
$n_{0}=10^{15}$~cm$^{-3}$, and
$N_{0d}kT=2\times10^{9}~\mathrm{cm}^{-2}$ }\vskip3mm
\end{figure}

\begin{figure}
\includegraphics[width=\column]{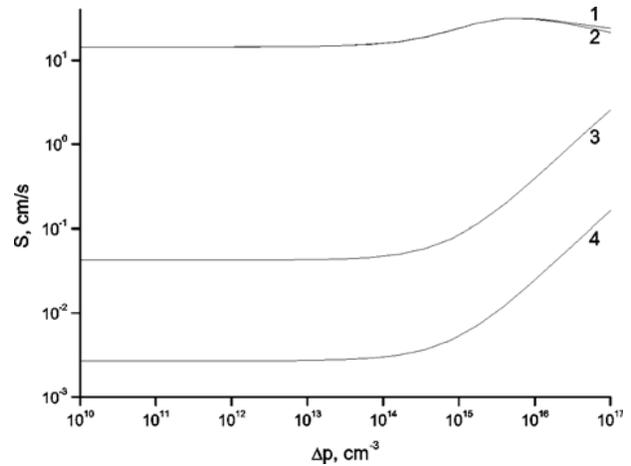}
\vskip-3mm\caption{Theoretical dependences of the effective surface
recombination rates $S_{s}$ (\textit{1}), $S_{d}$ (\textit{2}),
$S_{t}$ (\textit{3}), and $S_{a}$ (\textit{4}) on the injection
level. The calculation parameters are
$N_{\mathrm{ins}}=10^{12}~\mathrm{cm}^{-2}$,
$n_{0}=10^{15}$~cm$^{-3}$, and
$N_{0d}kT=2\times10^{9}~\mathrm{cm}^{-2}$  }\vspace*{-2mm}
\end{figure}

In the case $N_{\mathrm{ins}}=0$ (see Fig.~2), the effective rates
of surface recombination through the continuously distributed states
are low in comparison with that through the deep surface level. In
this case, the dependence of the effective surface recombination rates on
the injection level in the recession interval is approximated well
by the dependence $\Delta p^{-0.5}$, which is typical of
the recombination in the SCR \cite{7}. In the case where the
concentration of continuously distributed levels increases with the
energy more rapidly than the cross-sections of electron and
hole capture change, i.e. the inequality $E_{0}<E_{r}$ is obeyed,
and the integrated concentrations of continuously distributed levels
are comparable with the concentration of deep levels, the rates
$S_{a}$ and $S_{d}$ dominate.

\begin{figure}
\vskip1mm
\includegraphics[width=\column]{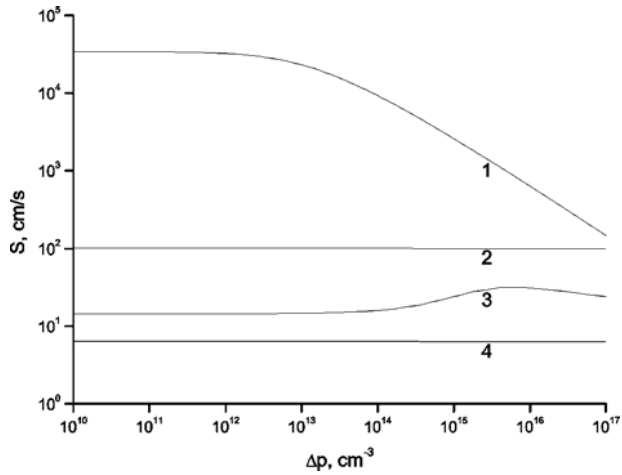}
\vskip-3mm\caption{Theoretical dependences of the effective surface
recombination rates $S_{s}$ (\textit{1}), $S_{a}$ (\textit{2}),
$S_{t}$ (\textit{3}), and $S_{da}$ (\textit{4}) on the injection
level. The calculation parameters are $N_{\mathrm{ins}}=0$,
$n_{0}=10^{19}$~cm$^{-3}$, and
$N_{0d}kT=2\times10^{9}~\mathrm{cm}^{-2}$  }
\end{figure}

In the case $N_{\mathrm{ins}}=10^{12}~\mathrm{cm}^{-2}$ (see
Fig.~3), the contribution of continuously distributed donor levels
to the surface recombination rate substantially exceeds that of the deep
discrete level, and the total surface recombination rate is much
higher than the value obtained above, when only the recombination
through the deep surface level was taken into account. The
dependences of the rates of surface recombination through the deep
discrete level and the continuously distributed acceptor levels on
the injection level magnitude are typical of those obtained in work
\cite{8} in the case of inversion band bending, but the same
dependence for the effective surface recombination rate through the
continuously distributed donor levels is not. In particular, the
quantity $S_{d}$ does not depend on the injection level firstly; then,
it increases a little and, afterward, starts to decrease.

It should be noted that, for the parameter set used while plotting Fig.~3,
the integrated concentrations of continuously distributed surface levels are
close to that of the deep surface level. At the same time, the effective rate of
surface recombination through the donor states considerably exceeds that
through the deep level, and the rate of recombination through the acceptor
states is much lower than that through the deep level. This can be explained
by the fact that, in this case, the Fermi level at the surface is located
considerably lower than the middle of the energy gap, and, as a result,
the recombination fluxes from the conduction and valence bands onto the donor
levels are rather close, although they are considerably different for the
deep level and the acceptor states.

In Fig. 4, the theoretical dependences of the effective surface
recombination rates on the injection level are exhibited for the case of
a heavily doped semiconductor, when the bulk concentration of electrons
amounts to $10^{19}$~\textrm{cm}$^{-3}$. In this case, there is a very small
depletion band bending at the semiconductor surface, $y_{s}(\Delta
n=0)\approx -0.01$; and the rate of recombination through the continuously
distributed acceptor levels dominates. The latter considerably exceeds the
rate of recombination through the deep level at low injection levels, but
diminishes as the injection level grows further.

Therefore, if the continuously distributed surface levels can
substantially affect the magnitudes of surface band bending and
photovoltage, when the integrated concentrations of those states
become comparable with $N_{\mathrm{ins}}$, then the effective rates
of surface recombination depending on them can considerably exceed
the rate of surface recombination through the deep level at much
lower integrated concentrations, when they are comparable with the
deep level concentration.

The growth of the effective surface recombination rate $S_{d}$ with $N_{0d}$
results in a reduction of the injection level under open-circuit
conditions, which should indirectly affect the voltage in the open circuit,
by reducing it. To determine the injection level (the excess concentration)
under open-circuit conditions in silicon SCs with the
inversion channel, we must solve the equation of balance between generation
and recombination. In the case of thin (in comparison with the diffusion
length) SCs, this equation looks like
\begin{equation}
J_{g}=(S_{0}(\Delta p)+S_{d}(\Delta p)+d/\tau _{\mathrm{eff}}(\Delta
p))\Delta p,
\end{equation}%
where $J_{g}$ is the generations flux, $S_{0}$ the effective surface
recombination rate at the illuminated front surface of SC, $S_{d}$
its value at the rear surface, $d$ the SC thickness, and $\tau
_{\mathrm{eff}}$ the effective time of bulk recombination with regard for
both linear and nonlinear mechanisms of bulk
recombination \mbox{in silicon \cite{4}.}

\begin{figure}
\vskip1mm
\includegraphics[width=\column]{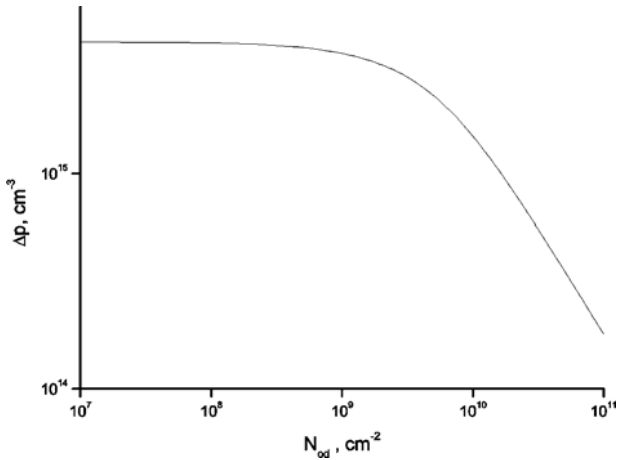}
\vskip-3mm\caption{Theoretical dependence of the injection level
under AM 1.5 conditions on the quantity $N_{0d}$. The calculation
parameters are $S_{d}=3$~cm/s,
$N_{\mathrm{ins}}=10^{12}~\mathrm{cm}^{-2}$,
$n_{0}=10^{15}$~\textrm{cm}$^{-3}$, $d=100 $~$\mu\mathrm{m}$, and
$\tau=10^{-3}$~s  }
\end{figure}

Figures~5 and 6 demonstrate the dependences of $\Delta p$ magnitude
in inverted silicon SCs under AM 1.5 conditions and the open-circuit
voltage $V_{\rm OC}$, respectively, on the concentration of
continuously distributed surface centers $N_{0d}$. The bottom
abscissa axis in Fig.~6 serves to indicate the corresponding
injection levels. Curve~\textit{1} corresponds to the case
$N_{0d}kT=10^{9}~\mathrm{cm}^{-2}$, curve \textit{3} to the case
$N_{0d}kT=10^{10}~\mathrm{cm}^{-2}$, and curve~\textit{2} was
plotted for $N_{0d}kT$ varying from $10^{9}$ to
$10^{10}~\mathrm{cm}^{-2}$. In the latter case, the exponentially
distributed surface centers practically do not influence the
magnitude of $V_{\rm OC}$. As one can see from Fig.~6, when the
quantity $N_{0d}kT$ reaches a value of $10^{10}~\mathrm{cm}^{-2}$,
the magnitude of $\Delta p^{\ast }$ decreases rather strongly,
whereas $V_{\rm OC}$ diminishes by approximately~10\%.

In Fig.~7, the dependences of the short-circuit current density
$J_{\rm SC}$ in silicon SCs with a metallized rear surface and under
AM 1.5 conditions on the concentration of continuously distributed
surface centers $N_{0d}$ are shown for the case where there is the
conductivity inversion at the SC front surface owing to the charge
built-in into the insulator. As one can see from the figure, a
reduction of $J_{\rm SC}$ as $N_{d}$ grows, which is connected with
the increase in the effective rate of surface recombination at the
front surface, is rather substantial and can reach 17\%, provided
that the integrated concentration of continuously distributed
centers and the concentration of deep levels are equal to each other
(at $N_{0d}kT=$ $=3\times 10^{10}~\mathrm{cm}^{-2}$). Hence, the
influence of continuously distributed surface centers on the
parameters of inverted silicon SCs owing to both the screening and
the enhancement of surface recombination can be \mbox{strong
enough.}\looseness=1

\begin{figure}
\vskip1mm
\includegraphics[width=\column]{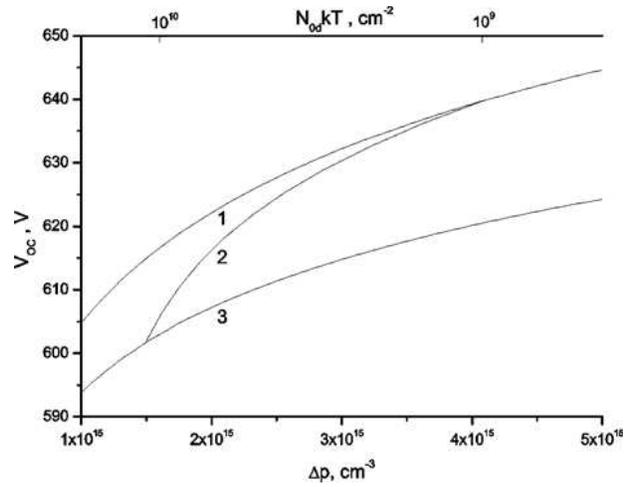}
\vskip-3mm\caption{Theoretical dependences of the open-circuit
voltage $V_{\rm OC}$ on $N_{0d}$ (the top abscissa axis) and the
corresponding injection level (the bottom abscissa axis) for
inverted silicon SCs under AM 1.5 conditions. The calculation
parameters are the same as in Fig.~5  }\vskip3mm
\end{figure}

\begin{figure}
\includegraphics[width=\column]{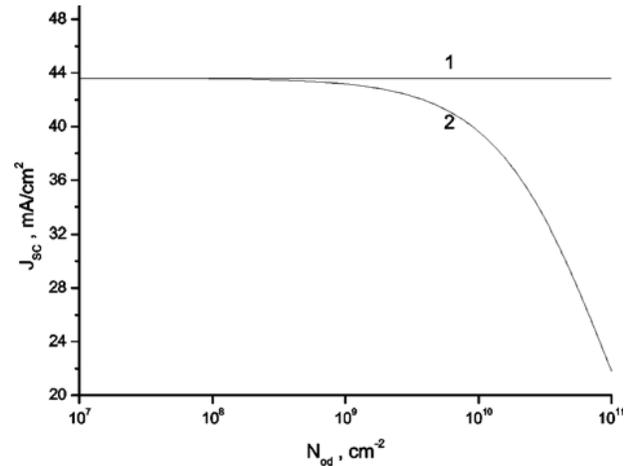}
\vskip-3mm\caption{Theoretical dependence of the short-circuit
current density in silicon SCs with a rear metallization on $N_{0d}$
under AM~1.5 conditions. The calculation parameters are $N_{\rm
ins}=$ $=10^{12}~\mathrm{cm}^{-2}$, $n_{0}=10^{15}
$~\textrm{cm}$^{-3}$, $d=100$~$\mu\mathrm{m}$, and $\tau=10^{-3}$~s
}
\end{figure}

While comparing the results presented in Fig.~7 for the dependence of
the short-circuit current density in the SC with the metallized rear surface and
the results of numerical 2D simulations described in work \cite{6} (see
Fig.~6), it becomes evident that they agree not only qualitatively, but also
quantitatively. The matter is that a reduction of the phototransformation
efficiency in SCs with rear metallization occurs mainly owing to the growth
of the surface recombination rate at the illuminated surface. Using the
calculation parameters given in work \cite{5}, one can easily get convinced
that the increase in the concentration $N_{d}(E)$ of exponentially
distributed donor surface centers within the limits shown in Fig.~7
corresponds to the growth of the surface recombination rate in the interval from
1 to 10$^{4}\mathrm{~cm/s}$ exhibited in Fig.~6 of work \cite{5}. The
corresponding reduction in the short-circuit current density completely
correlates with the reduction of the phototransformation efficiency in SCs
with rear metallization with respect to both the amplitude and the reduction
rate.

\begin{figure}
\vskip1mm
\includegraphics[width=\column]{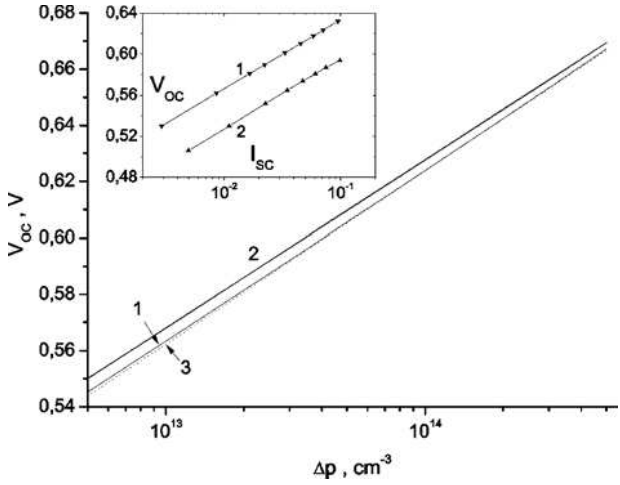}
\vskip-3mm\caption{Theoretical dependences of the open-circuit
voltage $V_{\rm OC}$ on the short-circuit current calculated by the
formulas of work \cite{9} (curves~\textit{1} and \textit{2}) and by
expression (12) (curve~\textit{3}). The experimental (points) and
theoretical (curves) dependences $V_{\rm OC}(I_{\rm SC})$ for the
SCs with the base of $n$- or $p$-types are shown in the inset  }
\end{figure}

Now, let us estimate the influence of exponentially distributed
surface centers in the case of diffusion silicon SCs. As the results
shown in Fig.~4 testify, the effective rate of surface recombination
in a heavily doped layer substantially increases owing to the
continuously distributed acceptor states. In diffusion SCs, this
configuration corresponds to the base of the $p$-type. As was shown,
e.g., in work \cite{8}, the effective $~\mathrm{s}$surface
recombination rate decreases in this case owing to both a stronger
doping of the $n$-type region created near the front surface and as
a result of its confinement by the rate of charge carrier delivery
to the heavily doped layer. The dependences of the open-circuit voltage
$V_{\rm OC}$ on the injection level can be calculated with the use of
expressions given in work \cite{9}. In Fig.~8, the corresponding
plots are exhibited for CSs with the $p$-type base. Curve~\textit{1}
was plotted with regard for the recombination through the
continuously distributed acceptor surface levels, and
curve~\textit{2} involves only the surface recombination
through the deep level located in the middle of the energy gap.
Curve~\textit{3} is described by the expression
\begin{equation}
V_{\rm OC}(\Delta p)=0.0259 A \ln(\Delta p/p_{0}).
\end{equation}%
Here, $A$ is the factor of CVC non-ideality, and $p_{0}$ the
equilibrium concentration of holes in the base bulk. As one can see
from curves~\textit{1} and \textit{3}, they are in quite a good
agreement if $A=1.12$ and $p=10^{7}$~\textrm{cm}$^{-3}$.

The inset in Fig.~8 demonstrates the experimental dependences
$V_{\rm OC}(I_{\rm SC}) $ for two specimens of diffusion silicon SCs
with the base of $n$- or $p$-type. For the SC with the base of
$n$-type, the values $V_{\rm OC}(I_{\rm SC})$ are larger than those for
the SC with the base of $p$-type. The factors of CVC non-ideality
for those SCs amount to 1.12 and 1.16, respectively, which is
evident from the dependences calculated by the formula
\begin{equation}
V_{\rm OC}(I_{\rm SC})=0.0259 A \ln(I_{\rm SC}/I_{s}),
\end{equation}%
where $I_{s}$ is the saturation current (curves~\textit{1} and \textit{2},
respectively). A theoretical fitting showed that, for diffusion SCs with the
base of $n$-type, the factor of non-ideality  equals 1.12, and the saturation
current is $3.4\times 10^{-11}~\mathrm{A}$. For the specimen with the base
of $p$-type, $A=1.12$ and $I_{s}=2.4\times 10^{-10}~\mathrm{A}$. If we take
into consideration that the saturation current $I_{s}=qp_{0}S_{s}$, provided
that the rate of surface recombination considerably exceeds that of
recombination in the bulk, and put $S_{s}=2.1\times 10^{4}$~cm/s and
$p_{0}=10^{4}$~\textrm{cm}$^{-3}$, we obtain $I_{s}\approx 3.4\times
10^{-11}~\mathrm{A}$.

Let us rewrite expression (12) in the form
\begin{equation*}
V_{\rm OC}(I_{\rm SC})\!=\!0.0259 A \ln\!\left(\!\frac{q\Delta
pS_{s}}{qp_{0}S_{s}}\!\right)\!\equiv\!
 0.0259 A \ln\!\left(\!\frac{I_{\rm SC}}{I_{s}}\!\right)\!.
\end{equation*}
Then it becomes evident that dependences (12) and (13) are
equivalent in the case $A=1.12$ and $S=$
$=2.1\times10^{4}~\mathrm{cm/s}$. This means that the theoretical
dependence described by curve~\textit{2} in Fig.~8 agrees with the
experimental dependence $V_{\rm OC}(I_{\rm SC})$ for a SC with the
base of $n$-type.

Note that the consideration of the surface recombination through continuously
distributed surface centers allows not only the
theoretical and experimental dependences $V_{\rm OC}(I_{\rm SC})$ to
be put in agreement, but also the difference of the CVC non-ideality
factor from 1 for diffusion SCs at high enough injection levels
to be explained.

\rezume{%
В.П.~Костильов, А.В.~Саченко,\\ І.О.~Соколовський, В.В.~Черненко,\\
Т.В.~Слусар, А.В.~Суший}{ВПЛИВ СИСТЕМИ ПОВЕРХНЕВИХ\\ ЦЕНТРІВ НА
ЕФЕКТИВНУ ШВИДКІСТЬ\\ ПОВЕРХНЕВОЇ РЕКОМБІНАЦІЇ\\ ТА НА ПАРАМЕТРИ
КРЕМНІЄВИХ\\ СОНЯЧНИХ ЕЛЕМЕНТІВ} {В роботі наведено результати
дослідження впливу рекомбінаційних експоненціально розподілених
поверхневих центрів на процеси поверхневої рекомбінації та на
характеристики кремнієвих сонячних елементів (СЕ). При розрахунках
вважалося, що система акцепторних поверхневих станів розташована в
верхній половині забороненої зони кремнію, а донорних поверхневих
станів~-- в нижній. Враховувалось також, що біля середини
забороненої зони знаходиться дискретний поверхневий рівень.
Показано, що у випадку, коли інтегральна концентрація неперервно
розподілених центрів порівняна з концентрацією глибокого
поверхневого рівня, вони впливають на характеристики кремнієвих СЕ
лише завдяки рекомбінації. У другому випадку, коли їх концентрація
порядку чи більша за концентрацію, яка характеризує вбудований в
діелектрику заряд, вони безпосередньо впливають на величину
поверхневого вигину зон та на значення фото-ерс. В результаті
комп’ютерного моделювання встановлено умови, за яких швидкість
рекомбінації через неперервно розподілені поверхневі центри більша
за швидкість рекомбінації через глибокий дискретний рівень.
Розраховано зменшення напруги розімкненого кола в кремнієвих
інверсійних СЕ, пов'язане з рекомбінацією через неперервно
розподілені центри. Розвинуту теорію порівняно з експериментом.}

\end{document}